\documentclass[10pt,letterpaper]{article}
\usepackage[top=0.85in,left=2.75in,footskip=0.75in]{geometry}

\usepackage{amsmath,amssymb}

\usepackage{changepage}

\usepackage[utf8x]{inputenc}

\usepackage{textcomp,marvosym}

\usepackage{cite}

\usepackage{nameref,hyperref}

\usepackage[right]{lineno}

\usepackage{microtype}
\DisableLigatures[f]{encoding = *, family = * }

\usepackage[table]{xcolor}

\usepackage{array}

\newcolumntype{+}{!{\vrule width 2pt}}

\newlength\savedwidth

\usepackage{setspace} 
\raggedright
\usepackage[aboveskip=1pt,labelfont=bf,labelsep=period,justification=raggedright,singlelinecheck=off]{caption}

\makeatletter
\renewcommand{\@biblabel}[1]{\quad#1.}
\makeatother

\usepackage{lastpage,fancyhdr,graphicx}
\usepackage{epstopdf}
\fancyheadoffset[L]{2.25in}
\fancyfootoffset[L]{2.25in}
\usepackage{color}
\usepackage{soul}

\renewcommand{\hl}[1]{#1}

\newcommand{\be}{\begin{equation}}
\newcommand{\ee}{\end{equation}}
\newcommand{\bra}{\langle}
\newcommand{\ket}{\rangle}
\newcommand{\bea}{\begin{eqnarray}}
\newcommand{\eea}{\end{eqnarray}}

\begin{document}
\vspace*{0.2in}

\begin{flushleft}
{\Large
\textbf\newline{Time-varying properties of asymmetric volatility and multifractality in Bitcoin} 
}
\newline
\\
Tetsuya Takaishi\textsuperscript{1*},
\\
\bigskip
\textbf{1} Hiroshima University of Economics, Hiroshima, Japan
\bigskip

* tt-taka@hue.ac.jp

\end{flushleft}
\section*{Abstract}
This study investigates the volatility of daily Bitcoin returns
and multifractal properties of the Bitcoin market
by employing the rolling window method and examines relationships
between the volatility asymmetry and market efficiency.
Whilst we find an inverted asymmetry in the volatility of Bitcoin,
its magnitude changes over time, and recently, it has become small.
This asymmetric pattern of volatility also exists in higher frequency returns.
Other measurements, such as kurtosis, skewness, average, serial correlation, and multifractal degree, also change over time.
Thus, we argue that properties of the Bitcoin market are mostly time dependent. 
We examine efficiency-related measures: the Hurst exponent, multifractal degree, and kurtosis.
We find that when these measures represent that the market is more efficient,
the volatility asymmetry weakens.
For the recent Bitcoin market, both efficiency-related measures and the volatility asymmetry
prove that the market becomes more efficient.


\section{Introduction}

Bitcoin, advocated by Satoshi Nakamoto\cite{Nakamoto2008}, was launched in 2009 as the first decentralized cryptocurrency. Its system is based on a peer-to-peer network. 
Whilst many other cryptocurrencies have been created since its launch, and the cryptocurrency market has grown rapidly, 
Bitcoin remains the dominant cryptocurrency in terms of market capitalization. 
Fig 1 represents the market capitalizations of the largest 10 cryptocurrencies. Bitcoin dominates about 70\% of the total capitalization of 10 cryptocurrencies.

\vspace{10mm}
\begin{figure}[h]
\vspace{5mm}
\centering
\includegraphics[width=9cm]{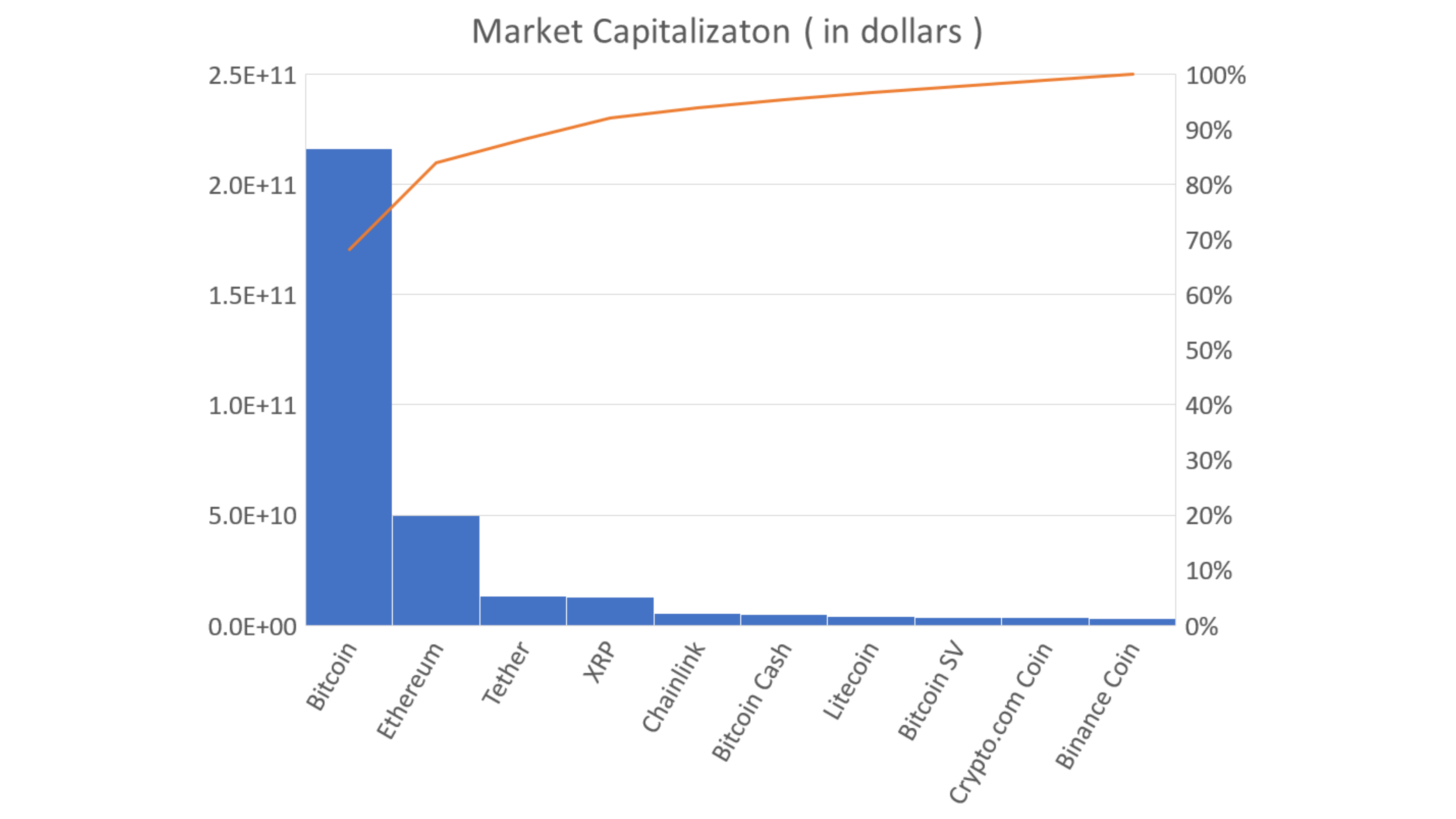}
\caption{
Market capitalization of the largest 10 cryptocurrencies (as of 31 August).
The solid line represents a Pareto chart.
The data are taken from "https://coinmarketcap.com/."
}
\end{figure}
\vspace{10mm}

In recent years, Bitcoin has attracted interest of many researchers.
Various aspects of Bitcoin, including 
hedging capabilities\cite{dyhrberg2016hedging}, bubbles\cite{cheah2015speculative},
liquidity and efficiency\cite{wei2018liquidity}, Taylor effect\cite{takaishi2018taylor},
structural breaks \cite{thies2018bayesian},
transaction activity\cite{koutmos2018bitcoin},
complexity synchronization\cite{fang2018multiscale},
long memory effects\cite{phillip2019long},
price clustering\cite{urquhart2017price}, rough volatility\cite{takaishi2019rough}
power-law cross-correlation\cite{Takaishi_2020}, market structure\mbox{\cite{drozdz2020complexity}}
have been investigated.

Similar to other assets, stylized facts\mbox{\cite{Cont2001QF,kwapien2012physical}}, such as volatility clustering, fat-tailed return distribution, and long memory in absolute returns, 
are observed in Bitcoin (e.g., \cite{chu2015statistical,takaishi2018}). 
An aggregational Gaussianity that the fat-tailed return distributions change to the Gausian distribution on large time scales is  another stylized fact
observed in various assets\mbox{\cite{Cont2001QF,kwapien2012physical}}.
This aggregational Gaussianity is also observed in Bitcoin and a minimum time scale required to recover
the Gaussianity is estimated to be two weeks\mbox{\cite{takaishi2018}}. 
The skewness is negative on a short time scale, but it moves to zero on a larger time scale\cite{takaishi2018}. 

Despite the similar stylized facts in Bitcoin, researchers report 
a distinct property: inverted volatility asymmetry. 
By using generalized autoregressive conditional heteroscedasticity (GARCH) models\cite{Engle1982autoregressive,Bollerslev1986JOE,Nelson1991Econ,Glosten1993JOF},
several studies report an "inverted asymmetry" in which volatility reacts more to
positive returns than negative ones\cite{bouri2016return,katsiampa2017volatility,stavroyiannis2017dynamic,ardia2019regime}. 
This contrasts sharply with the observation that the volatility of stocks reacts to negative returns more than positive ones\cite{Black1976,Christie1982stochastic,wu2001determinants}. 

However, some studies report that there is no significant asymmetry in volatility \cite{dyhrberg2016bitcoin,takaishi2018}.
An opposite result to the inverted asymmetry, that is, the same volatility reaction as stocks, is documented\cite{bouoiyour2016bitcoin}.
Moreover, whilst inverted asymmetry is also observed in other cryptocurrencies\cite{baur2018asymmetric,fakhfekh2020volatility},
the asymmetry in Bitcoin is insignificant\cite{baur2018asymmetric}.
Therefore, a consistent picture of volatility asymmetry in Bitcoin has not been obtained.
We infer that the discrepancy observed in the volatility asymmetry is caused, in part, by 
the time-varying property of volatility asymmetry. We attribute the differing conclusions to the various data periods used in earlier studies.
The possibility of time-varying asymmetry has been already noted\cite{bouri2016return}, and
it has been reported that whilst inverted asymmetry is observed before 2014, 
no significant asymmetry is observed after 2014. 

This study aims to investigate time-varying properties of the Bitcoin market, 
especially volatility asymmetry.  
In emprical finance, one of popular models to analyze volatility is the GARCH model\cite{Engle1982autoregressive,Bollerslev1986JOE} which can successfully capture 
some stylized facts such as volatility clustring and fat-tailed distribution.
There exist many variants of the GARCH model designed to capture more properties of financial time series\mbox{\cite{bollerslev1992arch}}.
To assess the volatility asymmetry, we use the threshold GARCH (TGARCH) model\cite{Glosten1993JOF},
which is widely accepted in empirical finance and has been used in the previous studies on 
the Bitcoin volatility asymmetry\cite{bouri2016return,bouoiyour2016bitcoin,baur2018asymmetric,takaishi2018,ardia2019regime,kyriazis2019estimating,fakhfekh2020volatility}.
We estimate model parameters by employing the rolling window method, which enables us to see time variation.
The rolling window method is commonly used in econometrics or empirical finance for time-series analysis in a limited amount of financial data.
We also investigate the Hurst exponent and multifractality of the Bitcoin market by the multifractal detrended fluctuation analysis (MF-DFA) method\cite{kantelhardt2002multifractal}, 
which is a powerful method to study multifractal properties and has been applied for various assets in econophysics; see, for example, \cite{Jiang-Xie-Zhou-Sornette-2019-RPP}.
The Hurst exponent and multifractality of the Bitcoin time series 
are examined intensively in connection with the market efficiency (e.g., \mbox{\cite{urquhart2016inefficiency,bariviera2017some,bariviera2017inefficiency,alvarez2018long,kristoufek2018bitcoin,takaishi2018,da2018multifractal,zhang2018inefficiency,takaishi2019market,dimitrova2019some}}). 
Here note that according to the efficient market hypothesis\cite{Fama1970efficient}, there are three types of market efficiencies: weak, semi-strong, and strong forms.
Since we use the time-series data only, the market efficiency in this study means the weak-form market efficiency.

A remarkable feature that the Bitcoin time series exhibits is presence of anti-persistency, that is, the Hurst exponent less than 1/2 \cite{urquhart2016inefficiency}.
The anti-persistency means that the time series reverses its moving direction more often than a random time series.
On the other hand, the time series with the Hurst exponent greater than 1/2 persists the same moving direction more than a random time series.
This anti-persistency behavior, however, turns out to be temporary. 
It is observed that the Hurst exponent and the multifractality degree vary over time, and the anti-persistency appears repeatedly\cite{alvarez2018long,takaishi2019market}.
Then, it seems that the Hurst exponent approaches the value of 0.5, which might be an indication toward a maturity market\mbox{\cite{urquhart2016inefficiency,drozdz2018bitcoin}}.

Since the efficient market should be free from any type of volatility asymmetry that results in predicting a certain market property to help gaining profits, 
the volatility asymmetry is expected to induce some inefficiency and to relate with
efficiency-related measures such as multifractality.   
Thus, we also examine a possible relationship between volatility asymmetry and efficiency-related measures.
We find that efficiency-related measures are related to the volatility asymmetry. 
When the efficiency-related measures indicate that the market is more efficient,
the volatility asymmetry weakens.                

To fully understand the dynamics of Bitcoin time series,
we need to investigate various aspects of Bitcoin.
This study investigates not only the volatility asymmetry but also
the multifractality, and combines them to advance the understanding of properties of the Bitcoin market.
Our results reveal that the Bitcoin market efficiency has improved in recent years.
It has been claimed that the market efficiency is related with market size and economic development\cite{zunino2012efficiency}. 
In accordance with this, our results suggest that the Bitcoin market is more mature than ever.

The rest of this paper is organized as follows. Section 2 describes the methodology. 
 In Section 3, we describe the data, and in Section 4, we present the empirical results and discuss the results.
 Finally, we conclude our study in Section 5.

\section{Methodology}

Let $p_{t_i}; t_i= i\Delta t;  i=0,1,2,...,N$ be the Bitcoin price time series
with sampling period $\Delta t$.
We define the return $r_{t_{i+1}}$ by the logarithmic price difference:
\be
r_{t_{i+1}}=100\times (\log p_{t_{i+1}} -\log p_{t_{i}}).
\ee
Since in empirical finance the volatility analysis using the GARCH-type models mainly focuses on daily volatility, 
here we also focus on daily returns, i.e. $\Delta t=1440-min$.

The TGARCH model\cite{Glosten1993JOF} is used to investigate volatility asymmetry in the Bitcoin market; 
the return $r_{t_i}$ and the volatility $\sigma^2_{t_i}$ at ${t_i}$ are modelled as follows:
\bea
r_{t_i} & = &\mu + c_1 r_{t_{i-1}} + \epsilon_{t_{i}}, \\
\sigma^2_{t_i} & = & \omega + \alpha \epsilon_{t_{i-1}} +\beta \sigma^2_{t_{i-1}} + \gamma \epsilon_{t_{i-1}}^2 I(\epsilon_{t_{i-1}}),
\eea
where $\epsilon_{t_i}$ is defined by $\epsilon_{t_i} = \sigma_{t_i} \eta_{t_i}$, and
$I(\epsilon_{t_{i-1}})$ is an indicator function, implying that it is 1 if $\epsilon_{t_{i-1}}<0$ and 0 otherwise.
$\eta_{t_i}$ is an unobservable random variable from an independent and identically distributed (IID) process.
Here, we use the Student t distribution as an IID process.
To check the robustness on choice of distributions,
we use the normal distribution and the generalized error distribution\cite{Nelson1991Econ}.

It is empirically well-known that stock return volatility increases after negative returns more than positive returns\cite{Black1976,Christie1982stochastic}.
This volatility asymmetry is called "the leverage effect" and causes a negative correlation between stock returns and volatility.
To capture the leverage effect, various GARCH-type models with the volatility asymmetry are introduced, e.g. \mbox{\cite{Glosten1993JOF,Nelson1991Econ,Sentana1995RES,heston2000closed,takaishi2017rational}}.
For the TGARCH model, the volatility asymmetry is measured by the $\gamma$ parameter in Eq (3),
and when the leverage effect exists, the $\gamma$ parameter takes a positive value.
On the other hand, for the inverted volatility asymmetry observed in the Bitcoin market the $\gamma$ parameter takes a negative value and volatility reacts more to positive returns than negative ones, leading to the inverted volatility asymmetry.
$c_1$ is the coefficient of an autoregressive model of order 1 (AR(1)) that captures the serial correlation. 

To investigate the time-varying properties of volatility asymmetry,
we use the rolling window method to estimate the model parameters. For the parameter estimation, we use the "urgarch" package of R.
First, we set a window size of 548 days ($\approx$ one and a half years). We choose this window size because the smaller window size such as one year 
leads to more noisy results in estimating model parameters,
and perform a parameter estimation for the window containing the first 548 data samples for the time series.
Next, we shift the window 30 days ($\approx$ one month) 
and perform a parameter estimation for the data in the next window.
We repeat this process until the end the time series.

Multifractal analysis is a useful method to quantify properties of complex system,  and it has been applied in many different fields, e.g.\mbox{\cite{ivanov1999multifractality,stanley1988multifractal,lennartz2008long,ihlen2013multifractal,jafari2007long}}.
Multifractal analysis is also popular in studies of financial markets and multifractal properties have been intensively studied\mbox{\cite{jafari2007long}}.
To investigate the multifractal properties of the Bitcoin market, we apply the MF-DFA method\cite{kantelhardt2002multifractal}.
The MF-DFA method is described by the following steps.

(i) Determine the profile $Y(i)$.
\be
Y(i)=\sum_{j=1}^i (r_{t_j}- \bra r \ket),
\ee
where $\bra r \ket$ stands for the average of returns.

(ii) Divide the profile $Y(i)$ into $N_s$ non-overlapping segments of equal length $s$, where $N_s \equiv {int} (N/s)$.
Since the length of the time series is not always a multiple of $s$, a short time period at the end of the profile may remain.
For this part, the same procedure is repeated starting from the end of the profile.
Therefore, total $2N_s$ segments are obtained.

(iii) Calculate the variance.
\be
F^2(\nu,s)=\frac1s\sum_{i=1}^s (Y[(\nu-1)s+i] -P_\nu (i))^2,
\ee
for each segment $\nu, \nu=1,...,N_s$ and
\be
F^2(\nu,s)=\frac1s\sum_{i=1}^s (Y[N-(\nu-N_s)s+i] -P_\nu (i))^2,
\ee
for each segment $\nu, \nu=N_s+1,...,2N_s$.
Here, $P_\nu (i)$ is the fitting polynomial to remove the local trend in segment $\nu$;
we use a cubic order polynomial.

(iv) Average over all segments and obtain the $q$th order fluctuation function.
\be
F_q(s)=\left\{\frac1{2N_s} \sum_{\nu=1}^{2N_s} (F^2(\nu,s))^{q/2}\right\}^{1/q}.
\label{eq:FL}
\ee
For $q=0$, the averaging procedure in Eq(\ref{eq:FL}) cannot be directly applied.
Instead, we employ the following logarithmic averaging procedure.
\be
F_0(s)=\exp\left[ \frac1{4N_s} \sum_{\nu=1}^{2N_s} \ln (F^2(\nu,s))\right].
\ee

(v) Determine the scaling behavior of the fluctuation function.
If the time series $r_{t_i}$ are long-range power-law correlated,
$F_q(s)$ is expected to be the following functional form for large $s$:
\be
F_q(s) \sim s^{h(q)}.
\label{eq:asympto}
\ee
In calculating the fluctuation function $F_q(s)$, we take $q$ varying between -25 and 25, with a step of 0.2.
The scaling exponent $h(q)$ is called the generalized Hurst exponent, and
the usual Hurst exponent is given by $h(2)$. 
When $h(q)$ is constant, the time series is called "monofractal." 
For example, the random Gaussian time series is monofractal\cite{kantelhardt2002multifractal}. 
On the other hand, when $h(q)$ varies depending on $q$, 
the time series is called "multifractal."

Following \cite{zunino2008multifractal}, we define the multifractality degree \hl{$\Delta h(q)$} by
\be
\Delta h(q) = h(q_{min})-h(q_{max}),
\ee
\hl{where $q_{min}=-q$ and $q_{max}=q$.}

Since for the random Gaussian time series, \hl{$\Delta h(q)$} takes zero, 
the magnitude of \hl{$\Delta h(q)$} is expected to relate with the strength of the market inefficiency.

We also calculate the singularity spectrum $f(\alpha)$, which is another way to characterize a multifractal time series. It is defined by
\be
\alpha (q) = h(q)+qh^\prime (q),
\ee
\be
f(\alpha)=q[\alpha -h(q)]+1,
\ee
where $\alpha(q)$ is the H\"older exponent or singularity strength\cite{kantelhardt2002multifractal}.
The multifractality degree through $f(\alpha)$ is defined by
\be
\Delta \alpha(q) = \alpha(q_{min})- \alpha(q_{max}),
\ee
which 
takes zero for monofractal.
In this study we take $q=[-25,25]$. However, there is a caution that when $|q|$ is large, the moments may diverse and the estimated results are unreliable\mbox{\cite{Jiang-Xie-Zhou-Sornette-2019-RPP}}.
Thus, to avoid such a situation, following \mbox{\cite{drozdz2018bitcoin}},  for the calculations of the multifractal degree \hl{$\Delta h(q)$ and $\Delta \alpha(q)$, we take $q=4$.}
The strength of \hl{$\Delta \alpha (q)$} is also related with the market inefficiency, and we call \hl{$\Delta h(q)$} and \hl{$\Delta \alpha(q)$} including kurtosis
"efficiency-related measures."
We use the same rolling window method to investigate time-varying properties of multifractality
and take the same window size (548 days) with the volatility analysis by the TGARCH model to compare the results.

\section{Data}
We use Bitcoin tick data (in dollars) traded on Bitstamp
from September 10, 2011, to
June 06, 2020, downloaded from Bitcoincharts (http://api.bitcoincharts.com/v1/csv/).
Due to a hacking incident, no data are available from January 4, 2015 to January 9, 2015. For these missing data,
we treat them as the price is unchanged.

\subsection{Price and return}
Fig 2 illustrates daily ($\Delta t=1440$-min) price and returns $r_{t_j}$ constructed from the Bitcoin tick data. We eliminate the data that are larger than 40, i.e., $r_{t_i}>40$, as outliers.
This manipulation keeps the results almost unchanged except kurtosis. In our data set, we find four outliers. 
Fig 2(d) shows the volatility series $s_t$ defined by 
$s_t = s_{t-1} + |r_t| -\bar{r}$, where $r_t$ and $\bar{r}$ stand for the return at $t$ and the average of $r_t$, respectively. 
The volatility series introduced in \mbox{\cite{segovia2019novel,nikolova2020novel}} can be utilized to identify the volatility clustering.
Namely, the increasing (decreasing) trend of the volatility series implies the existence of the high (low) volatility clustering.
Such trends indicating high and low volatility clusterings are seen in Fig 2(d).

\vspace{5mm}
\begin{figure}[h]
\vspace{5mm}
\centering
\includegraphics[width=9cm]{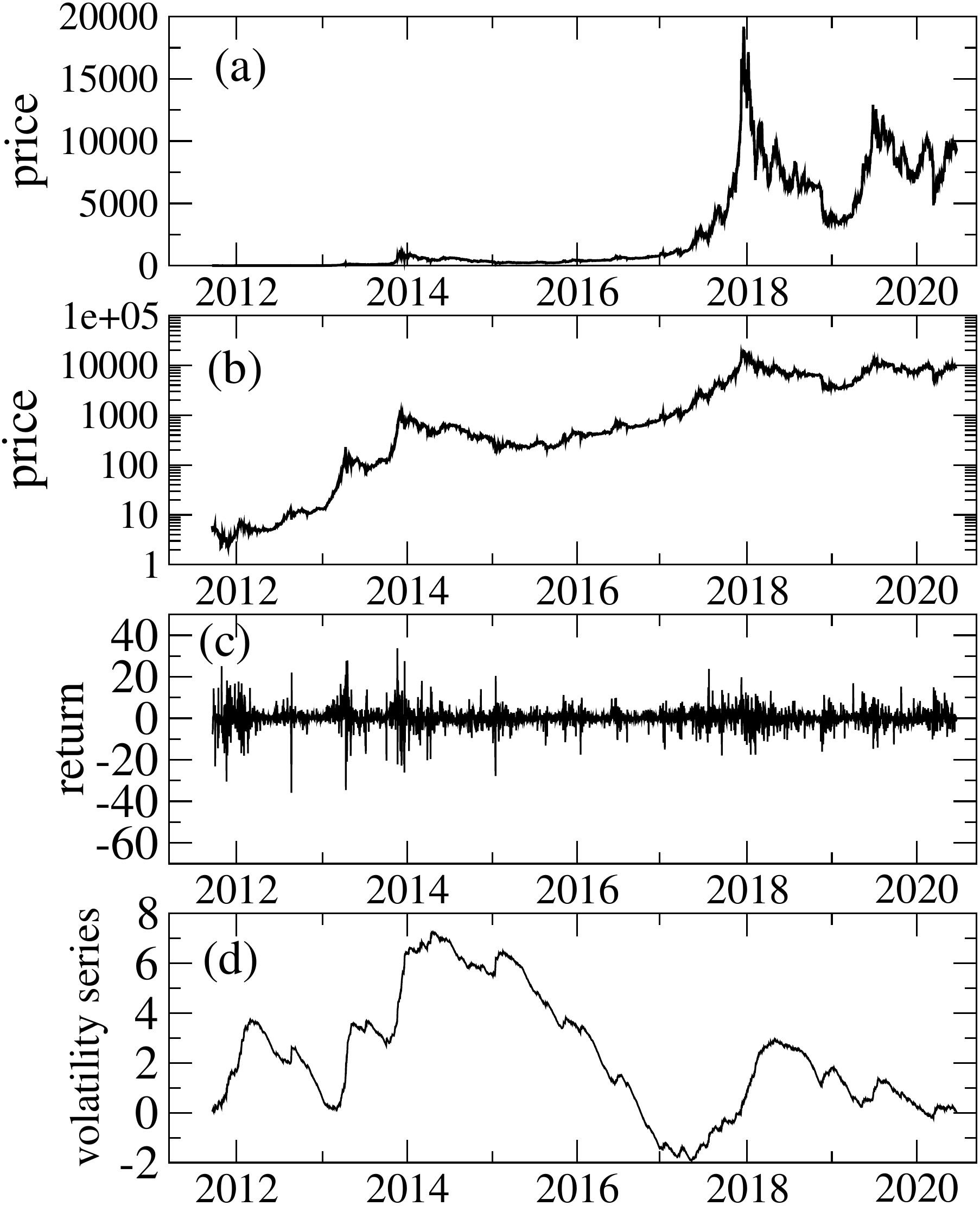}
\caption{
(a): Daily price, (b): daily price (semi-log plot), (c): returns and (d) volatility series.
}
\end{figure}
\vspace{5mm}

Table 1 provides descriptive statistics for the whole sample of returns,
and we find a positive average, high kurtosis, and negative skewness.
We also explore the time variation in these using the rolling window method.
\begin{table}[h]
\centering
\caption{
Descriptive statistics for the whole sample of daily returns. \\
SD stands for "standard deviation."
The values in parentheses indicate one sigma errors estimated by the Jackknife method.
}
\hspace{-5mm}
\begin{tabular}{ccccc}
\hline
   Mean & SD   & Kurtosis & Skewness & Nobs   \\
\hline
    0.27(11) & 4.63(41)  &  11.9(22) & -0.267(99) & 3188  \\
\hline
\end{tabular}
\end{table}

\subsection{Time-varying properties of descriptive statistics}
Fig 3 illustrates the average, standard deviation (SD), kurtosis, and skewness calculated with a 548-day rolling window. 
Interestingly, they vary considerably over time.  
Whilst the kurtosis before 2017 is very high (i.e., more than 10), it decreases after 2017. Recently, it has taken a value of around 6, which is still
higher than the Gaussian kurtosis. 
The origin of high kurtosis could be a fat-tailed return distribution that means higher price variations are observed more often.
At the early stage of Bitcoin market, the tail index $\mu$ of the cumulative return distribution is found to be $\mu \approx 2$,
which is referred to as the inverse square law\cite{easwaran2015bitcoin}. 
The similar tail indces are have also been reported in \mbox{\cite{beguvsic2018scaling}} 
This is sharply contrast to the well-known inverse cubic law for other assets\cite{gopikrishnan1998inverse,gopikrishnan1999scaling,plerou1999scaling,pan2007self}, 
in which the tail index $\mu$ is $\mu \approx 3$.
The inverse square law observed in the Bitcoin market, however, is not permanent.
The recent Bitcoin data up to 2017 show that the tail index comes close to 3, which suggests that
the Bitcoin market is becoming more mature\mbox{\cite{drozdz2018bitcoin}}.
Further studies\mbox{\cite{drozdz2019signatures,takaishi2020jpcs}} also indicate the change of the tail index to 3.
It is also worth noting that the recent COVID-19 pandemic considerably affects the cryptocurrency market 
and as a result the market experiences a volatile period in which the tail index varies\mbox{\cite{drozdz2020complexity}}.

These observations imply that the tail index $\mu$ of the cumulative return distribution in 
the Bitcoin market varies over time and moves from $\mu \approx 2$ (fatter tail) to $\mu \approx 3$ (thinner tail),
which could account for the decrease in kurtosis. 
Although the skewness is mostly negative over the whole period, 
the magnitude of the negative skewness decreases gradually with time, and 
the skewness seems to disappear in recent returns.
The disappearance of skewness possibly means the efficiency improvement of the Bitcoin market,
and agrees with the observation in efficiency-related measures such as the Hurst exponent and the multifractal degree that indicate that
the market efficiency of the Bitcoin market varies over time and the recent market efficiency is being improved. 
We describe the results on the market efficiency in more detail later.
Overall, these findings in descriptive statistics also suggest that properties of the Bitcoin market vary over time.
Thus, it is important to consider time variation in analysis.

\vspace{5mm}
\begin{figure}[h]
\centering
\includegraphics[width=9cm]{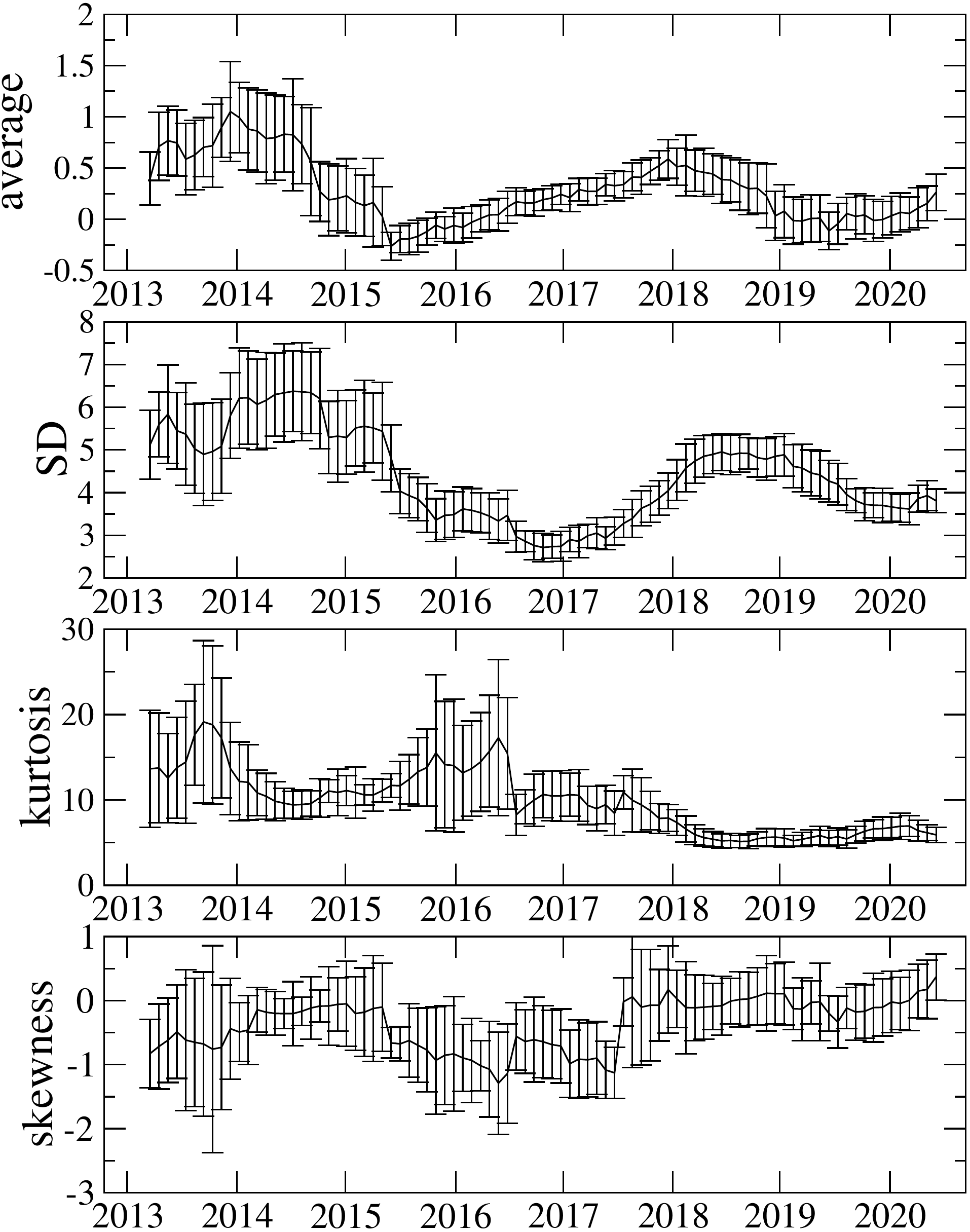}
\caption{
\bf Fig 3. Average, SD, kurtosis, and skewness as a function of time.
SD stands for standard deviation.
These are calculated using the rolling window method with a window size of 548 days.
Bars in the data points represent one sigma error, estimated by the Jackknife method.
}
\end{figure}
\vspace{5mm}

\subsection{Aggregational Gaussianity}
Fig 4 illustrates the aggregational Gaussianity of returns, namely the kurtosis of returns sampled at $\Delta t$ as a function of $\Delta t$.
The figure is plotted in log-log scale, and we find that the kurtosis decreases according to a power-law up to $\Delta t \approx 20000$ min $\approx$ two weeks, 
with an exponent $\sim -0.62$, and
the kurtosis of returns at $\Delta t$ longer than two weeks is consistent with the Gaussian kurtosis, in agreement with the previous result\cite{takaishi2018}.
This finding suggests that the time series of returns at $\Delta t$ longer than two weeks most likely becomes the random Gaussian time series.

\vspace{5mm}
\begin{figure}[h]
\vspace{5mm}
\centering
\includegraphics[width=9cm]{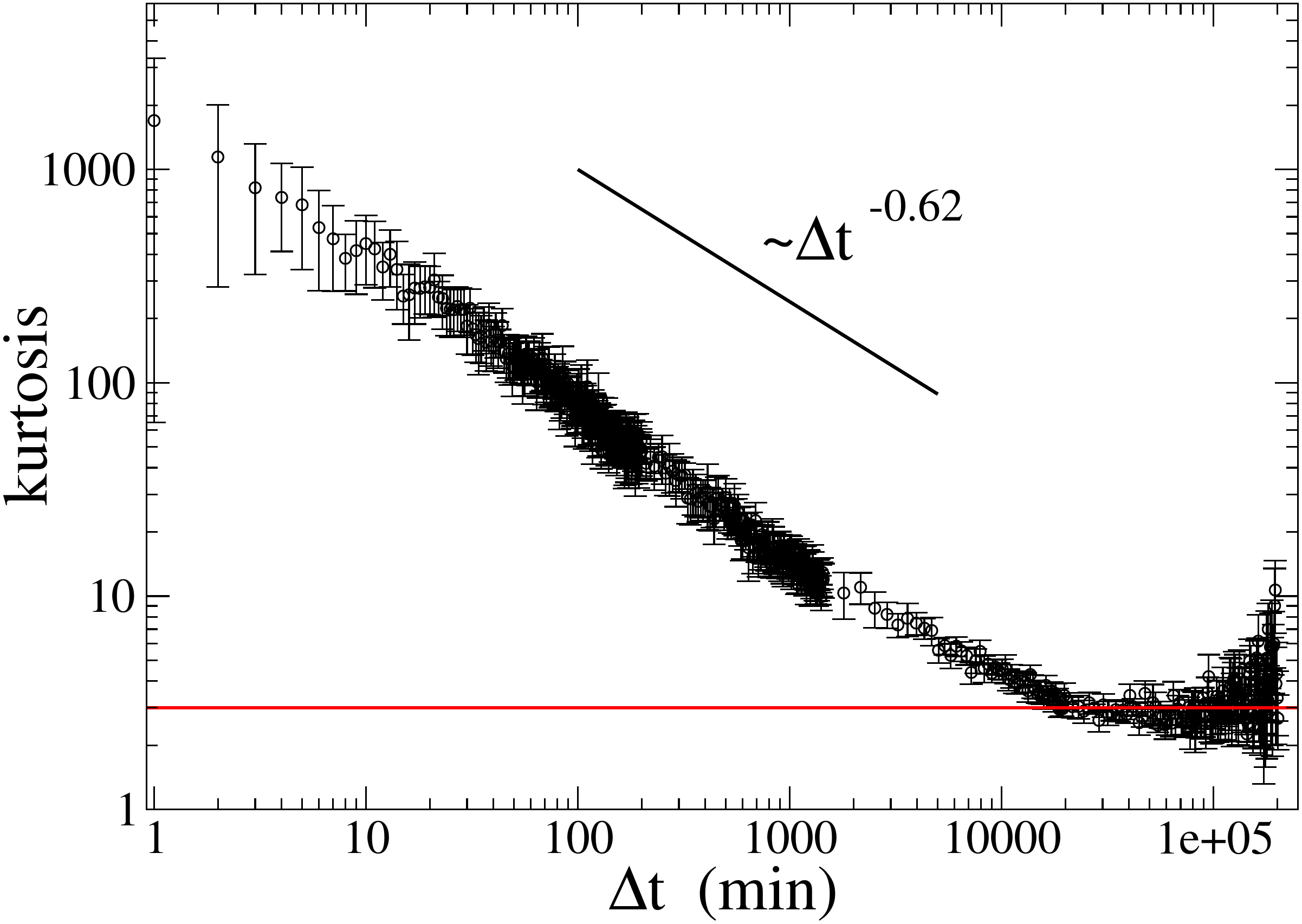}
\caption{
Kurtosis at various sampling period $\Delta t$ as a function of $\Delta t$.
Bars in the data points represent one sigma error, estimated by the Jackknife method.
The red line displays the Gaussian kurtosis ($=3$).
}
\end{figure}
\vspace{5mm}

\section{Empirical results and discussion}

Table 2 presents the TGARCH parameters estimated for the whole dataset. 
The $\gamma$ parameter is negative, which indicates an inverted asymmetry in the volatility.
Its magnitude, however, is not large, which is consistent with the result of \cite{baur2018asymmetric}.
This is probably because the strength of time-varying asymmetry of $\gamma$ weakens in the parameter estimation for the whole dataset
which may include both periods of positive and negative asymmetries. 
In the following, we find that the $\gamma$ parameter varies over time considerably.

\begin{table}[h]
\centering
\caption{
TGARCH parameter estimates for the whole dataset. $\nu$ is the shape parameter of the Student t distribution.\\
The values in parentheses indicate standard errors.
}
\hspace{-10mm}
\begin{tabular}{ccccccc}
\hline
   $\alpha$  & $\beta$  & $\omega$ & $\gamma$  & $c_1$ & $\mu$ &  $\nu$   \\
\hline
    0.181(36) & 0.853(17)  & 0.235(77) & -0.058(21) & -0.044(17)  & 0.181(36) & 3.21(14)  \\
\hline
\end{tabular}
\end{table}

Figs 5 and 6 show the results of the TGARCH parameters in the 548-day rolling window.
It is evident that the parameters vary over time.
Although parameter $\gamma$ is mostly negative, 
we find some exceptions. 
For instance, parameter $\gamma$ takes positive or zero values around 2015;
and after 2019, its magnitude becomes small or consistent with zero. 
We also observe a strong inverted asymmetry between 2016 and 2018. 
For the robustness check on the IID distribution in Eq (2), we perform 
the parameter estimation with the normal and generalized error distributions,
and find that the similar asymmetric volatility patterns to that from the Student t distribution are obtained.
Therefore, the choice of distributions in the IID process is irrelevant.

\vspace{5mm}
\begin{figure}[h]
\centering
\includegraphics[width=9cm]{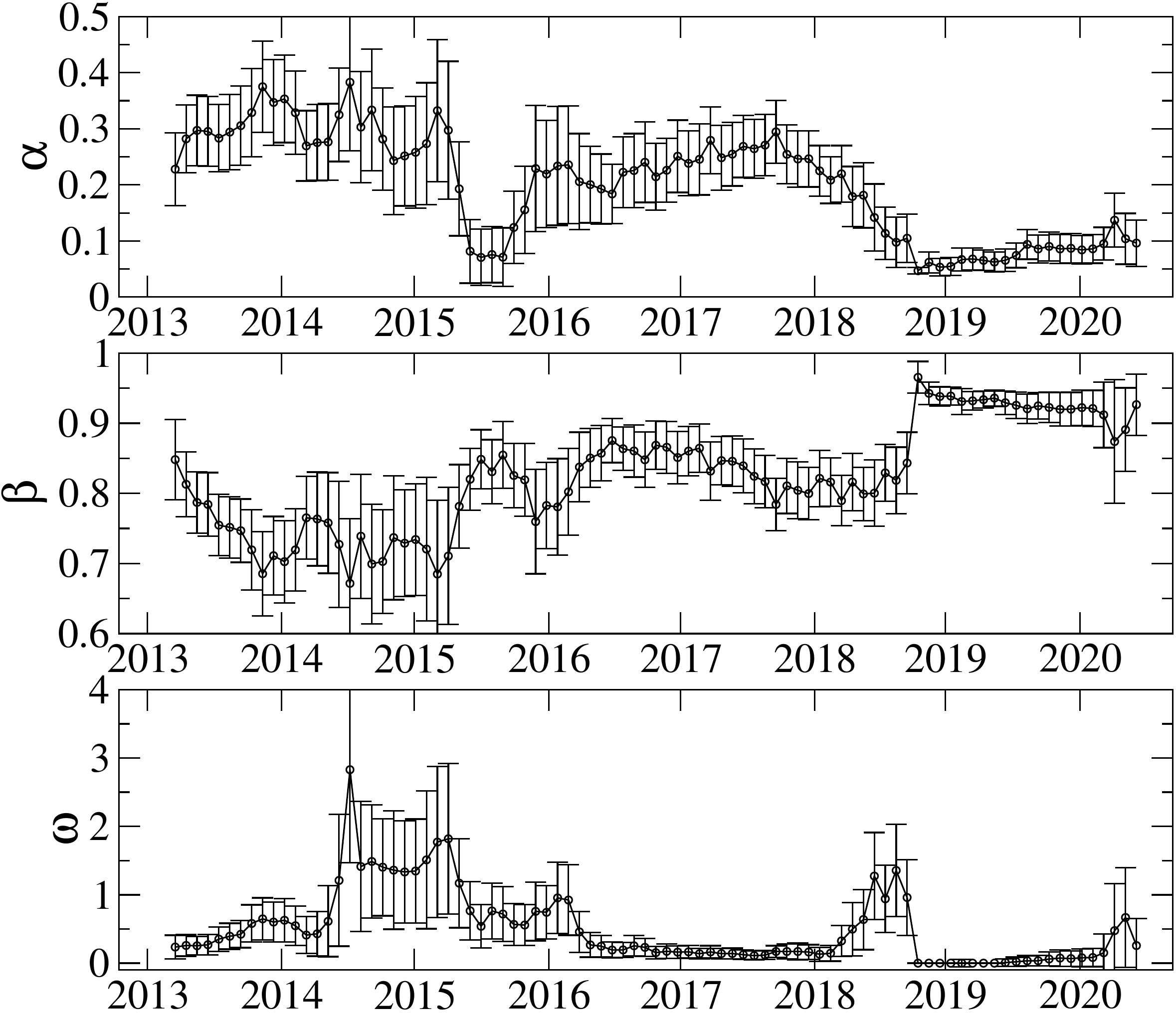}
\caption{
Estimation results of $\alpha,\beta$, and $\omega$.
The error bars show the standard errors.
}
\end{figure}
\vspace{5mm}

\vspace{5mm}
\begin{figure}[h]
\centering
\includegraphics[width=9cm]{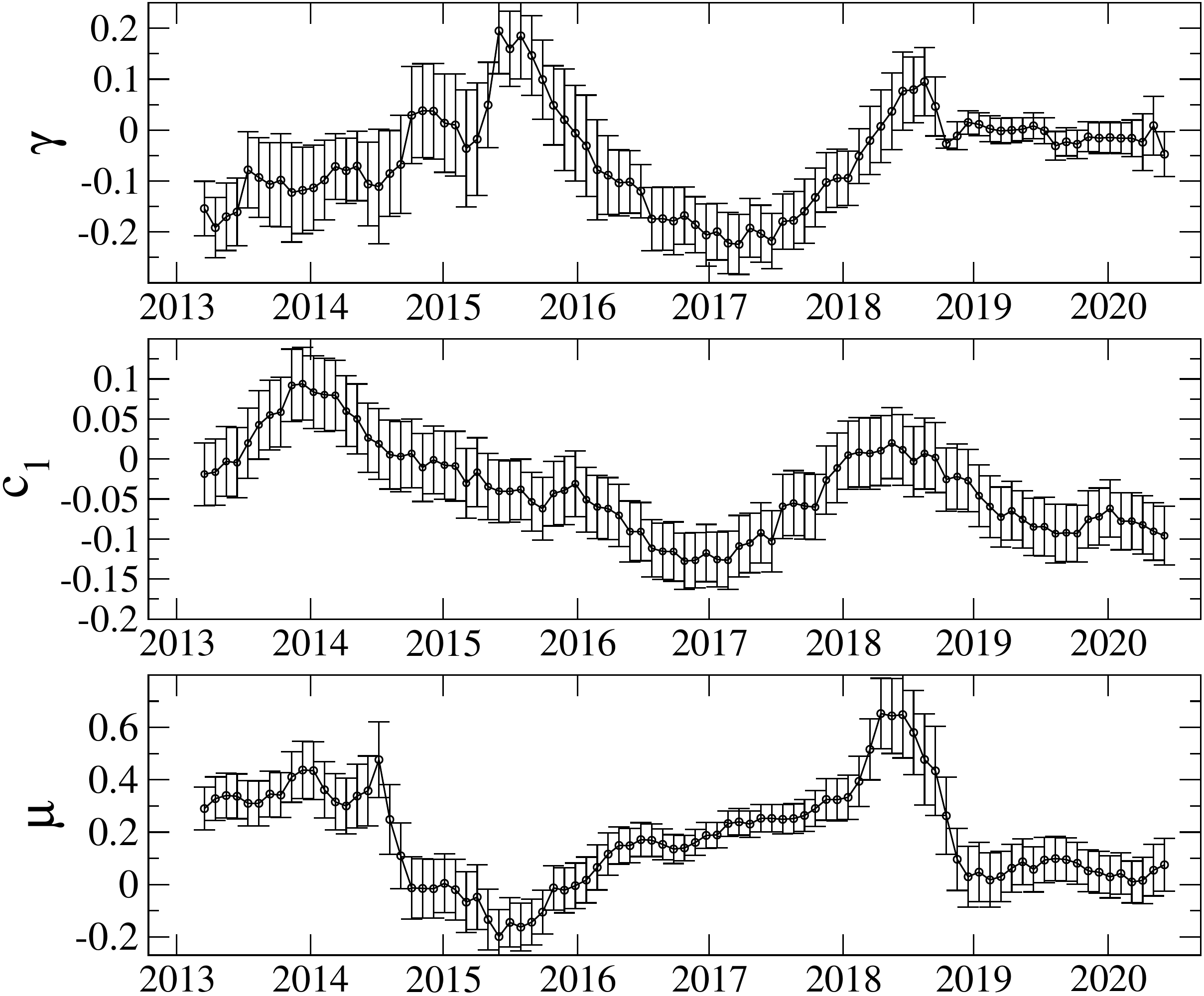}
\caption{
Estimation results of $\gamma,c_1$ and $\mu$.
The error bars show the standard errors.
}
\end{figure}
\vspace{15mm}

The AR(1) parameter $c_1$, which captures serial correlation, also varies considerably.
It is argued that non-zero serial correlation implies that uninformed investors dominate in trading 
and that price changes due to uninformed investors will increase volatility 
more than price changes caused by informed investors\cite{avramov2006impact}.
In line with \cite{avramov2006impact}, it is claimed that non-zero AR(1) coefficients are found for cryptocurrencies; 
and the inverted asymmetry due to uniformed investors is consistent 
with phenomena such as fear of missing out, pump and dump schemes, and the disposition effect\cite{baur2018asymmetric}.  
Our results of $c_1$ indicate that there are periods in which $c_1$ is consistent with zero,
which suggests that the Bitcoin market is not always dominated by uninformed investors.
We find both strong inverted asymmetry and non-zero $c_1$ from 2016 to 2018. Thus, 
the period from 2016 to 2018 is considered to be dominated by uninformed investors, thereby affecting the volatility asymmetry.
As seen in Fig 2, in this period, the Bitcoin price increases considerably and recorded the highest value on December 2017. 
Consequently, Bitcoin price movement in this period is associated with the inverted asymmetry induced dominantly by uninformed investors.

We also estimate the model parameters for high-frequency returns (6h and 12h).
Fig 7 represents the results of $\gamma$ together with those from the daily returns. 
We find that high-frequency returns exhibit similar variation with daily returns except that the 6h returns for which no significant inverted asymmetry is seen before 2015.
These results imply that whilst the asymmetric volatility pattern remains for higher frequency returns,
the detail of the asymmetry pattern depends on the frequency of returns.  

\vspace{15mm}
\begin{figure}[h]
\includegraphics[width=9cm]{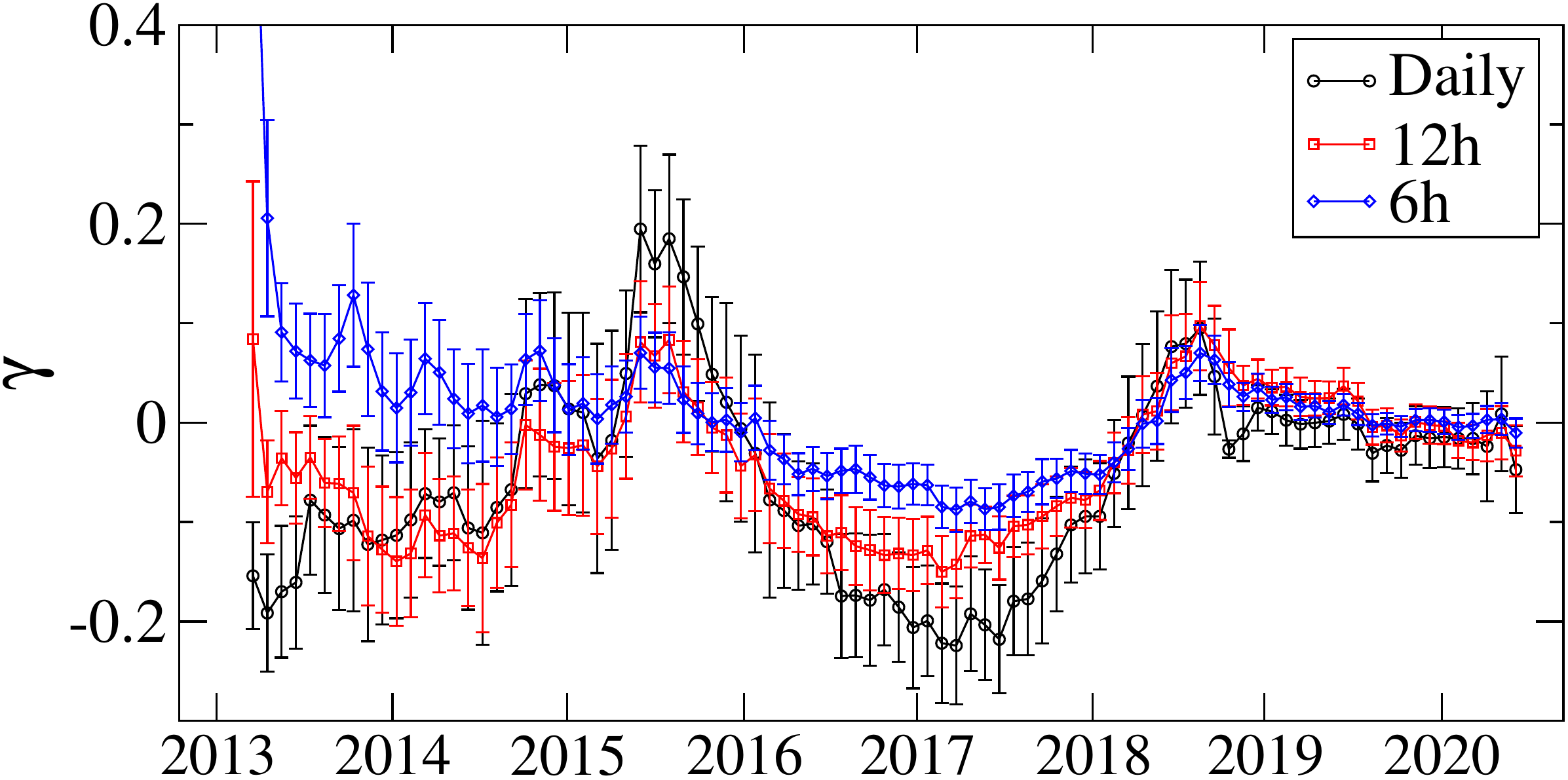}
\caption{
 Asymmetry parameter $\gamma$ for 6h, 12h, and daily returns.
The error bars show the standard errors.
}
\end{figure}
\vspace{5mm}

To calculate the multifractal degree \hl{$\Delta h(q)$ and $\Delta \alpha(q)$}, we first determine $h(q)$ by fitting the fluctuation function $F_q(s)$ to
Eq(9) in a range of $s=[20,100]$.
Then, we determine \hl{$\Delta h(q)$ and $\Delta \alpha(q)$} by Eq(10) and (13), respectively.
As representatives, Fig 8 (a)-(b) shows fluctuation functions $F_q(s)$ calculated using the first window data and $h(q)$, respectively.

\vspace{5mm}
\begin{figure}[h]
\centering
\includegraphics[width=9cm]{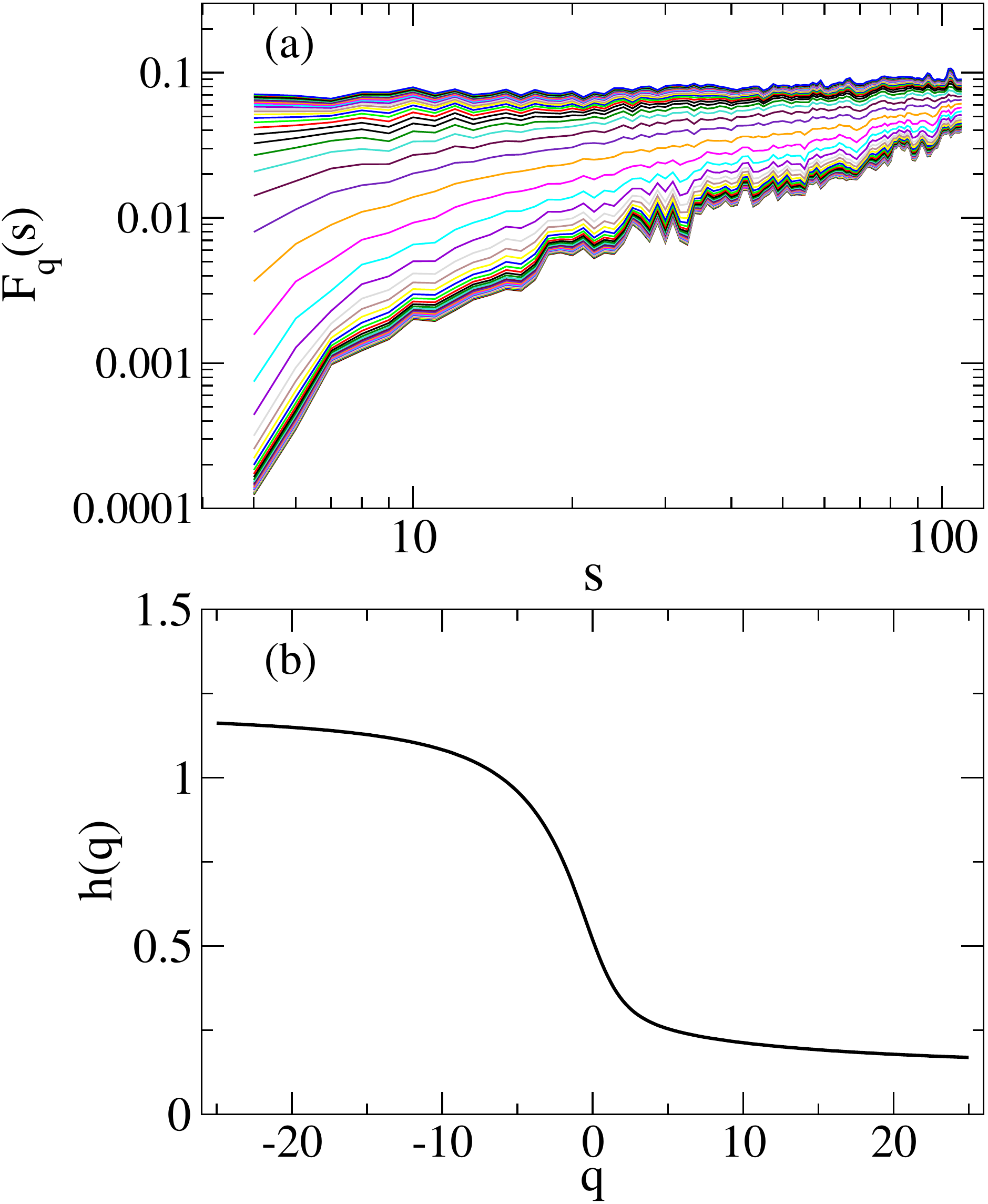}
\caption{
(a) Fluctuation functions $F_q(s)$ (b) The generalized Hurst exponent $h(q)$.
In Fig 8 (a), the results are plotted from $q=-25$ (bottom) to $q=25$ with a step of 1.0.
}
\end{figure}
\vspace{10mm}

Fig 9 (a) represents the Hurst exponent $h(2)$ as a function of time,
showing some anti-persistent periods ($h(2) <1/2$). The recent $h(2)$ around 2020 is consistent with or slightly above 1/2,
which suggests that the recent Bitcoin market is becoming more efficient.
This finding is consistent with the results of kurtosis and skewness in Fig 3 that shows low kurtosis and insignificant skewness for the recent Bitcoin market, meaning that 
the return distribution becomes more Gaussian shaped than before. 
For \hl{$\Delta h(q)$ and $\Delta \alpha(q)$}, we show the results of \hl{$q=4$} in Fig 9 (b) as a function of time.
Two measures of the multifractal degree calculated here show the same time-varying pattern. Broadly speaking, the multifractal degree decreases with time,
which suggests that the Bitcoin market gradually approaches the efficient market.
Using the Amihud illiquidity measure\cite{amihud2002illiquidity}, it is argued that the market inefficiency in the cryptocurrency market is caused by illiquidity and
the illiquidity is related with anti-persistency\cite{wei2018liquidity}.
For the Bitcoin market, liquidity in terms of the Amihud illiquidity measure turns out to be improving\cite{takaishi2019market}, which agrees with 
the improved efficiency of the Bitcoin market in recent years.

\vspace{25mm}
\begin{figure}[h]
\centering
\includegraphics[width=9cm]{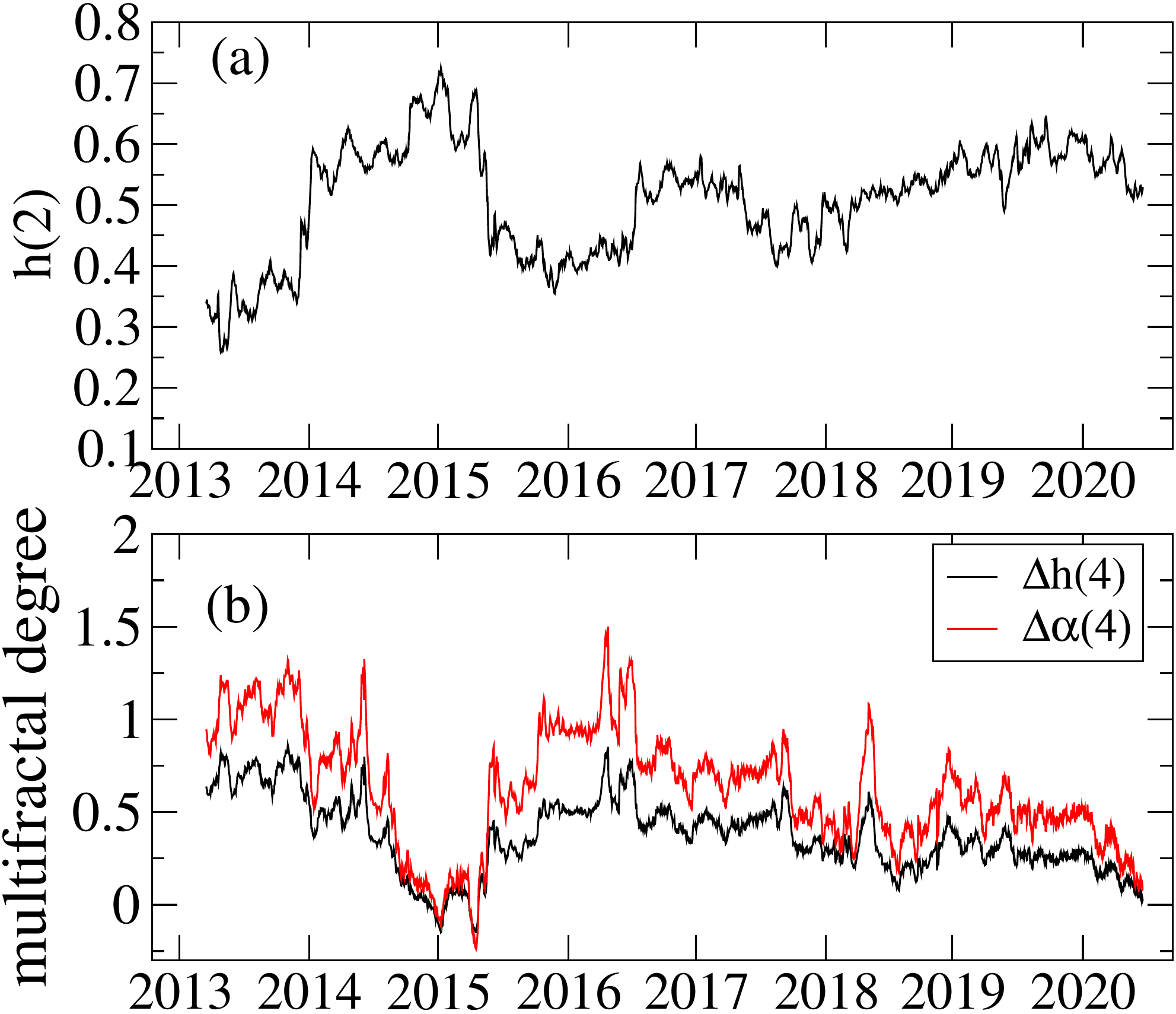}
\caption{
 (a) the Hurst exponent $h(2)$, (b) multifractal degrees $\Delta h(4)$ and $\Delta \alpha(4)$.
These results are obtained by the rolling window method with a 548-day window and a step of 1 day.
}
\end{figure}
\vspace{15mm}

In Fig 10 (a)-(b), we examine the relationship between kurtosis and the efficiency-related measures (a) $h(2)$ and (b) $\Delta \alpha(4)$.
Fig 10 (a) shows that the results of $h(2)$ near 1/2 takes kurtosis smaller values, for example, less than 7. 
However, most larger kurtoses more likely correspond to $h(2)$ far from 1/2.
In Fig 10 (b), we find that the smaller kurtoses opt to take small $\Delta \alpha(4)$.
Therefore, the results of Fig 10 (a)-(b) imply that
in the more efficient market, the kurtosis takes smaller values close to the Gaussian one.
Here, note that since returns at short scale are usually fat-tailed distributions, the kurtosis of daily returns may not reach the Gaussian one, and 
it rather reaches a certain minimum value ($>3$) at the efficiency-improved market.

\vspace{5mm}
\begin{figure}[h]
\centering
\includegraphics[width=9cm]{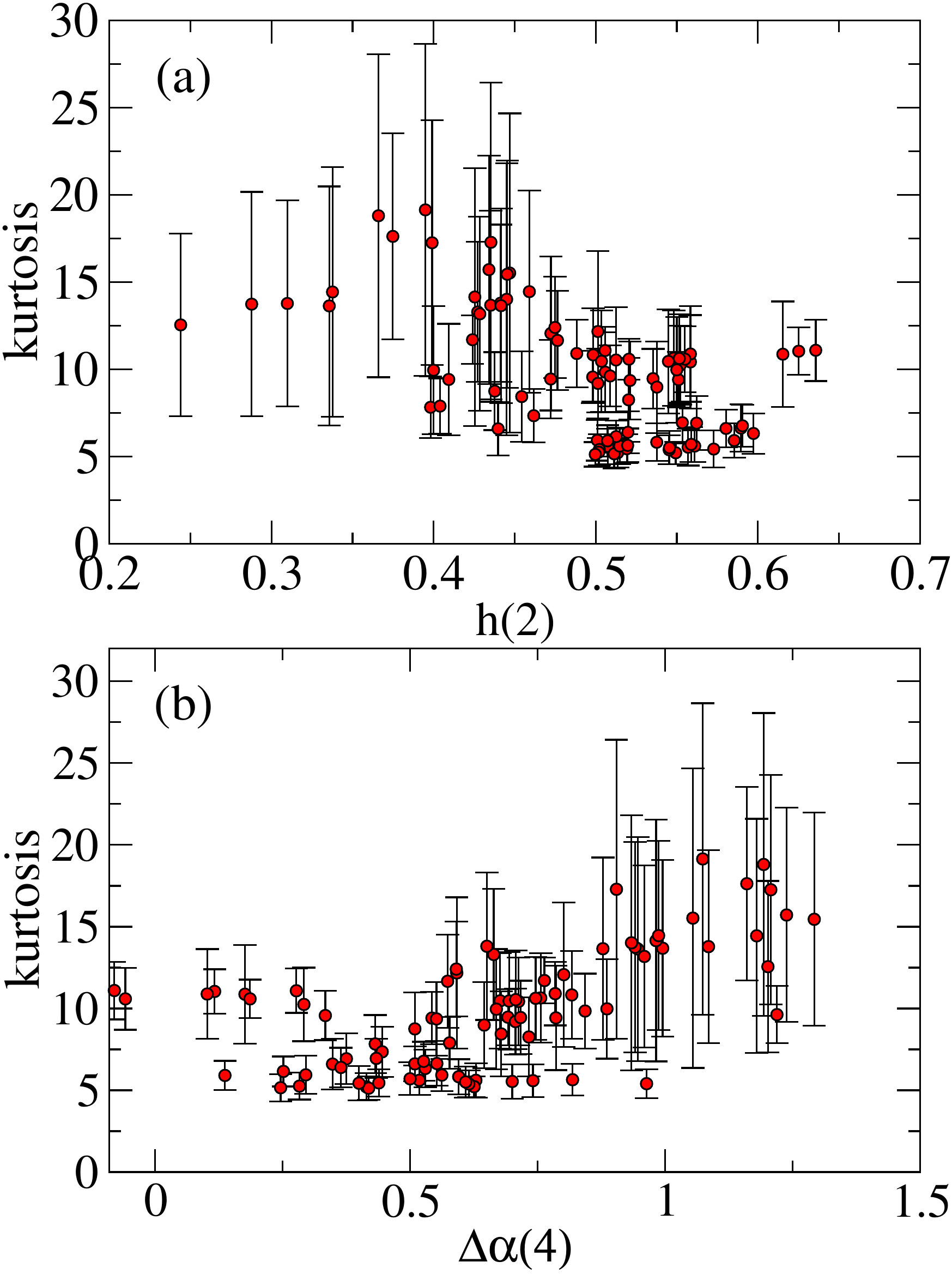}
\caption{
(a) Kurtosis versus $h(2)$ and (b) Kurtosis versus $\Delta \alpha(4)$.
}
\end{figure}
\vspace{5mm}

Fig 11 (a)-(c) illustrates relationships between the asymmetric parameter $\gamma$ and three measurements, (a) $h(2)$, (b) $\Delta \alpha(4)$, and (c) kurtosis,
and some remarks are in order.
In Fig 11 (a), it is observed that in an anti-persistent domain ($h(2)<1/2$), the asymmetric parameter $\gamma$ takes negative value more than positive ones,
and the parameter $\gamma$ near zero tends to cluster in a region that $h(2)$ takes values near 1/2 or slightly above 1/2. 

\vspace{5mm}
\begin{figure}[h]
\centering
\includegraphics[width=9cm]{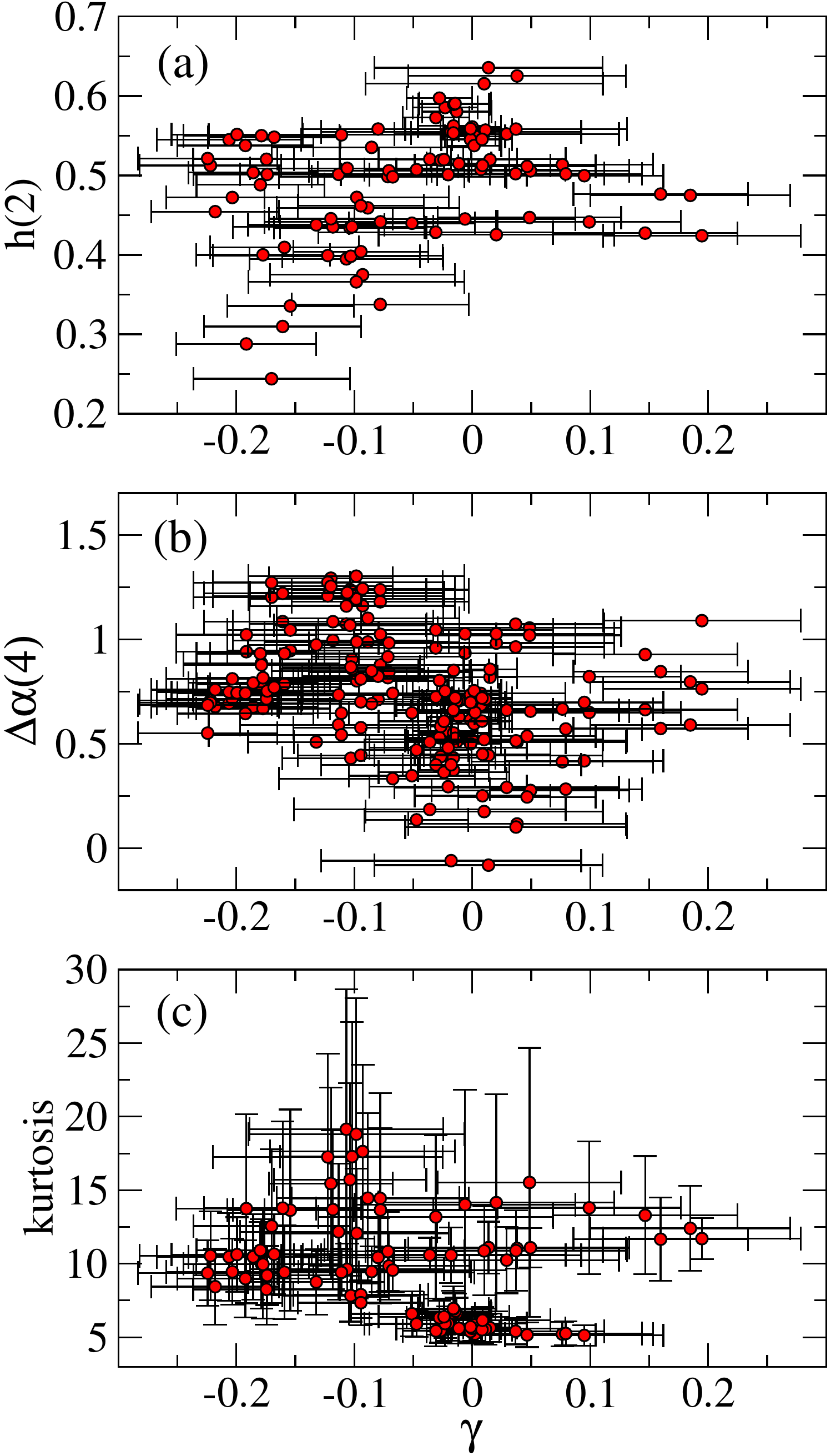}
\caption{
Relationships between $\gamma$ and (a) $h(2)$, (b) $\Delta \alpha(4)$, and (c) kurtosis.
}
\end{figure}
\vspace{10mm}
  
In Fig 11 (b), we find that in a region of small $\Delta \alpha(4)$, for example, $\Delta \alpha(4) < 0.5$, the parameter $\gamma$ mostly takes values near zero.
However, for larger $\Delta \alpha(4)$, the parameter $\gamma$ tends to take negative values.
Fig 11 (c) indicates that for a region near the Gaussian kurtosis (e.g., kurtosis $ < 7 $), the parameter $\gamma$ comes to values near zero.
Moreover, mostly the strong negative $\gamma$ comes to a region with higher kurtoses.
The overall results from Fig 10 (a)-(c) indicate that for the market being more efficient, that is, for $h(2)$ near 1/2, small $\Delta \alpha(4)$, and kurtosis closer to the Gaussian one,
the volatility asymmetry likely disappears. This is consistent with the consequence of the efficient market
that any predictable patterns such as the asymmetric volatility should not exist.

\subsection{An application on investment strategy}
Correct understanding of the market state is of great importance for investors
who change trading strategy depending on the state of the market.
This study could contribute to offer such information.
For example, according to the efficient market hypothesis\cite{Fama1970efficient}, the technical analysis is not supported on the efficient market.
Our results imply that the recent Bitcoin market is being efficient and it might be difficult to make high profits by the technical analysis. 
By monitoring the market efficiency of Bitcoin, if the Bitcoin market becomes inefficient substantially again, 
one could use the technical analysis to gain profits.
Another suggestion from the efficient market hypothesis is that on the efficient market, the most efficient portfolio is 
a market portfolio consisting of every asset weighted in proportion to its market capitalization. 
Of course, it is difficult to form a completely diversified portfolio in practice\cite{roll1977critique}.
However, it may be possible to make an approximate diversified portfolio including the efficient Bitcoin.
Furthermore, although not all cryptocurrencies are efficient,
when the cryptocurrency markets become more mature and more efficient,
an index or portfolio consisting of cryptocurrencies could be a proxy of a fully diversified portfolio on the cryptocurrency markets.

\section{Conclusions}
We use the rolling window method to investigate time-varying properties of Bitcoin. 
We find that various measurements, such as volatility asymmetry, kurtosis, skewness, serial correlation, and multifractality, are time varying.
Thus, the Bitcoin market may have inherently variable properties.  
Although the inverted asymmetry is observed in Bitcoin and the strong inverted asymmetry is found around 2016-2018,
the recent volatility asymmetry is weak.
The magnitude of the volatility asymmetry may relate with the market state, especially the market efficiency.

To investigate a relationship between the volatility asymmetry and the market efficiency,
we examine efficiency-related measures: the Hurst exponent, multifractal degree, and kurtosis.
We find that when these efficiency-related measures indicate that the market is more efficient,
the volatility asymmetry is more likely to weaken.
The efficiency-related measures indicate that the recent Bitcoin market has become more efficient.

Whilst we use one of the GARCH-type models, that is, TGARCH model, which is commonly used in volatility analysis of Bitcoin, 
other types of volatility models such as stochastic volatility model\cite{taylor1982,taylor1986modelling,asai2006multivariate} exist.
It might be interesting to investigate whether these models lead to the similar results on the volatility asymmetry. 

In addition to the inverted asymmetry, other remarkable properties are observed in the Bitcoin market.
For example, whilst it is claimed that for stock markets, time series of the log-volatility increments shows 
"monofractal" anti-persistence behavior\cite{gatheral2018volatility,bennedsen2016decoupling,livieri2018rough},
for the Bitcoin market "multifractal" anti-persistence behavior is reported\cite{takaishi2019rough}. This difference is important to construct a correct volatility model with monofractal or
multifractal behavior. 
Another interesting property is observed in return-volatility cross-correlation.
For stock markets, it is found that return-volatility cross-correlation function exhibits an exponential decay that indicates that
the return-volatility cross-correlation is short-ranged\cite{bouchaud2001leverage,qiu2006return,chen2013agent}. 
However, for the Bitcoin market, the return-volatility cross-correlation function shows a power law meaning that the cross-correlation is long-ranged\cite{Takaishi_2020}.
The mechanism that originates these remarkable properties including the inverted asymmetry are not established yet.
Toward the complete understanding of the Bitcoin dynamics, we should further study these properties in more detail in future direction.

\section*{Acknowledgment}
Numerical calculations for this work were carried out at the
Yukawa Institute Computer Facility and at the facilities of the Institute of Statistical Mathematics.



\end{document}